\documentclass[pdflatex,sn-mathphys-num]{sn-jnl}% Math and Physical Sciences Numbered Reference Style
\usepackage{graphicx}%
\usepackage{multirow}%
\usepackage{amsmath,amssymb,amsfonts}%
\usepackage{amsthm}%
\usepackage{mathrsfs}%
\usepackage[title]{appendix}%
\usepackage{xcolor}%
\usepackage{textcomp}%
\usepackage{manyfoot}%
\usepackage{booktabs}%
\usepackage{algorithm}%
\usepackage{algorithmicx}%
\usepackage{algpseudocode}%
\usepackage{listings}%
\usepackage{physics}
\usepackage{natbib}
\theoremstyle{thmstyleone}%
\theoremstyle{thmstyletwo}%

\theoremstyle{thmstylethree}%

\begin{document}

\title[Article Title]{Impact of Amplitude and Phase Damping Noise on Quantum Reinforcement Learning: Challenges and Opportunities}


\author*[1]{\fnm{Mar\'ia Laura} \sur{Olivera-Atencio}}\email{molivera1@us.es}

\author[2]{\fnm{Lucas} \sur{Lamata}}\email{llamata@us.es}

\author*[1,3]{\fnm{Jes\'us} \sur{Casado-Pascual}}\email{jcasado@us.es}

%\affil*[1]{\orgdiv{F\'{\i}sica Te\'orica}, \orgname{Universidad de Sevilla}, \orgaddress{\street{Street}, \city{Sevilla}, \postcode{Apartado de Correos 1065}, \state{State}, \country{Country}}}
\affil*[1]{\orgdiv{F\'{\i}sica Te\'orica}, \orgname{Universidad de Sevilla},  \postcode{Apartado de Correos 1065}, \city{Sevilla 41080}, \country{Spain}}

\affil[2]{\orgdiv{Departamento de F\'{\i}sica At\'omica, Molecular y Nuclear}, \orgname{Universidad de Sevilla}, \city{Sevilla 41080}, \country{Spain}}

\affil*[3]{\orgdiv{Multidisciplinary Unit for Energy Science}, \orgname{Universidad de Sevilla}, \city{Sevilla 41080}, \country{Spain}}


\abstract{Quantum machine learning (QML) is an emerging field with significant potential, yet it remains highly susceptible to noise, which poses a major challenge to its practical implementation. While various noise mitigation strategies have been proposed to enhance algorithmic performance, the impact of noise is not fully understood. In this work, we investigate the effects of amplitude and phase damping noise on a quantum reinforcement learning algorithm. Through analytical and numerical analysis, we assess how these noise sources influence the learning process and overall performance. Our findings contribute to a deeper understanding of the role of noise in quantum learning algorithms and suggest that, rather than being purely detrimental, unavoidable noise may present opportunities to enhance QML processes.}




\maketitle

\section{Introduction}\label{Introduction}
Quantum Machine Learning (QML) is a rapidly growing field within quantum technologies that seeks to perform machine learning tasks more efficiently than classical supercomputers in terms of time, space, and energy resources \cite{BiamonteNature, Schuld2021,MelnikovReview,LamataReview2,Wang_2024}. Leveraging quantum superposition and entanglement, the goal is to enable a more scalable implementation of various machine learning algorithms using quantum computers~\cite{NielsenChuang2000}.

A major challenge in quantum computing, which also affects QML, is the fragility of highly entangled many-body quantum states. These states are susceptible to interactions with unintended quantum systems, leading to decoherence and the loss of computational properties necessary to solve a given problem. However, an emerging perspective in QML explores decoherence and dissipation not only as an obstacle but also as a potential resource for enhancing quantum learning~\cite{Ghasemian2021,Ghasemian2022,Ghasemian2023,OliveraAtencio2023,Domingo2023,Sannia2024,OliveraAtencio2023_2}. This approach is motivated by the fact that effective learning, both classical and quantum, often requires some form of nonlinearity. Since isolated quantum systems evolve linearly (i.e., unitarily), some form of coupling---whether through quantum measurement (projective, weak, etc.) or dissipative and/or dephasing  processes governed by a master equation (Markovian or non-Markovian)---may play a crucial role in enabling richer learning dynamics. In this context, recent work has shown that carefully tuned amplitude, phase, and depolarizing noise can improve the performance of variational quantum algorithms, further supporting the idea that noise can be harnessed as a useful feature in QML~\cite{Somogyi2024}.

In previous works, we have explored the role of thermal dissipation in QML protocols~\cite{OliveraAtencio2023}. Our findings indicate that, rather than being purely detrimental, thermal dissipation can sometimes enhance the learning process. In this paper, we build upon this research by specifically analyzing phase damping noise (PDN) and amplitude damping noise (ADN)  in a protocol of quantum reinforcement learning.  By investigating these types of noise, we aim to deepen the understanding of when and how these types of noise can be beneficial in quantum learning protocols.

The remainder of this work is structured as follows. In Sec.~\ref{PS}, we provide a brief description of the problem under study, the types of noise considered, and the algorithm to which they are applied. In Sec.~\ref{results}, we present the numerical results and analyze the conditions under which noise can enhance the algorithm performance. Finally, in Sec.~\ref{conclusions}, we summarize our findings and conclusions.

\section{Problem Statement}\label{PS}

In the reinforcement learning algorithm presented in Ref.~\cite{Albarran_Arriagada_2020}, the agent A is a known and controllable quantum system described by a state vector $\ket{\phi}$, or equivalently, by the associated density operator $\rho=\ketbra{\phi}{\phi}$. For the sake of clarity, we will restrict ourselves to the simplest case, where the system is a single qubit with computational basis $ \{\ket{0}, \ket{1}\} $. The interaction of the environment E with the agent A over a time interval $ \tau $ is characterized by the unitary time evolution operator $U(\tau)=e^{-i H \tau/\hbar}$, where $H$ is an unknown Hamiltonian. This Hamiltonian, written in terms of its unknown excited state $\ket{e}$ and ground state $\ket{g}$, takes the form $H=\frac{\hbar \omega}{2} (\ketbra{e}{e}-\ketbra{g}{g})$, where $\omega$ is a characteristic frequency of the system.  The goal of the algorithm is to ``learn'' how to construct, at least approximately, either of the stationary states $\ket{e}$ or $\ket{g}$, despite the challenge posed by the fact that these stationary states are unknown. To this end, the algorithm exploits the fact that, for $\tau\neq \tau_n=2 n \pi/\omega$ with $n$ being a natural number, the stationary states $\ketbra{e}{e}$ and $\ketbra{g}{g}$ are the only pure states invariant under the unitary time evolution, i.e., the only pure states satisfying the equation  $U(\tau)\ketbra{\phi}{\phi}U^{\dagger}(\tau)=\ketbra{\phi}{\phi}$. Consequently, actions compatible with this property are rewarded, while those that are incompatible are penalized. The case $\tau=\tau_n$  is an exception since $U(\tau_n)=(-1)^n I$, with $I$ being the identity operator, making all pure states satisfy the condition $U(\tau_n)\ketbra{\phi}{\phi}U^{\dagger}(\tau_n)=\ketbra{\phi}{\phi}$.  For a detailed explanation of why this algorithm can be classified as a quantum reinforcement learning algorithm, see Refs.~\cite{OliveraAtencio2023,Albarran_Arriagada_2020}.

In the presence of noise, the time evolution that governs the interaction between the environment and the agent is no longer unitary. Specifically, for PDN and ADN---the cases analyzed in this work---the time evolution over an interval \(\tau\) takes the form 
\begin{equation}
\label{G_evolution}
\mathcal{E}_{\tau}(\rho)=U(\tau)\left[E_{0}(\tau)\rho E_{0}^{\dagger}(\tau)+ E_{1}(\tau)\rho E_{1}^{\dagger}(\tau)\right]U^{\dagger}(\tau),
\end{equation}
where $\{E_0(\tau),E_1(\tau)\}$ are Kraus operators~\cite{Kraus_1983} whose form depends on the type of noise considered~\cite{NielsenChuang2000}. Specifically, in the case of PDN, the Kraus operators and the corresponding time evolution take the form $E_0(\tau)=\ketbra{g}{g}+e^{-\tau/T_{\mathrm{D}}}\ketbra{e}{e}$, $E_1(\tau) =\sqrt{1-e^{-2 \tau/T_{\mathrm{D}}}}\ketbra{e}{e}$, and
\begin{equation}
\label{PD_evolution}
\mathcal{E}_{\tau}(\rho)=\rho_{ee}\ketbra{e}{e}+\rho_{gg}\ketbra{g}{g}
+e^{-\tau/T_{\mathrm{D}}}\left(e^{-i \omega \tau} \rho_{eg}\ketbra{e}{g}+e^{i \omega \tau}\rho_{ge}\ketbra{g}{e}\right),
\end{equation}
where $\rho_{\alpha \beta}=\mel{\alpha}{\rho}{\beta}$, with $\alpha,\beta\in\{e,g\}$, and $T_{\mathrm{D}}$ is a parameter known as the decoherence time~\cite{ZurekRMP03,BreuerPetruccione2003}. For ADN, the Kraus operator $E_0(\tau)$ remains the same as for PDN, while $E_1(\tau) =\sqrt{1-e^{-2 \tau/T_{\mathrm{D}}}}\ketbra{g}{e}$, leading to the time evolution  
\begin{equation}
	\begin{split}
		\label{AD_evolution}
		\mathcal{E}_{\tau}(\rho)=&\ketbra{g}{g}+\rho_{ee}e^{-2\tau/T_{\mathrm{D}}}(\ketbra{e}{e}-\ketbra{g}{g})\\
		&+e^{-\tau/T_{\mathrm{D}}}\left(e^{-i \omega \tau} \rho_{eg}\ketbra{e}{g}+e^{i \omega \tau}\rho_{ge}\ketbra{g}{e}\right).
	\end{split}
\end{equation}
In this case, in addition to the suppression of the off-diagonal elements of the density operator in the eigenstate basis of $H$ (decoherence), there is also a decay of the excited state $\ket{e}$ to the ground state  $\ket{g}$ with a mean decay time of $T_{\mathrm{D}}/2$. To some extent, this type of noise can be regarded as a specific instance of the thermal dissipation analyzed in Ref.~\cite{OliveraAtencio2023}, but at absolute temperature equal to zero. Note that, as in the unitary time evolution without noise, the stationary states $\ketbra{e}{e}$ and $\ketbra{g}{g}$ are the only pure states invariant under the non-unitary evolution induced by PDN [Eq.~(\ref{PD_evolution})].  In contrast, under ADN [Eq.~(\ref{AD_evolution})], only the ground state $\ketbra{g}{g}$ remains invariant. This asymmetry will be crucial to the algorithm performance in the presence of ADN, as discussed later. Next, we provide a detailed description of how the algorithm works in the presence of these types of noise.

The algorithm consists of a large number of iterations, indexed by a natural number  $k$. The unitary transformation generated in the $k$th iteration is denoted by $D_k$. Moreover, in each iteration, a numerical value within the interval $[0,1]$ is assigned to a parameter $w_k$, called the exploration parameter. The goal is for $D_k \ketbra{\phi}{\phi}D_k^{\dagger}$, with $\ketbra{\phi}{\phi}$ being a given state, to gradually approach either of the target states $\ketbra{e}{e}$ or $\ketbra{g}{g}$ as the number of iterations increases.  Starting with the initial values $D_0 = I$ and $w_0 = 1$, the values of $D_{k+1}$ and $w_{k+1}$ are iteratively updated from the previous values $D_k$  and $w_k$ according to the following steps:
\begin{enumerate}
\item The unitary transformation $D_k$ is applied to one of the computational basis state, say, the state $\ketbra{0}{0}$, to construct the state $\rho_k = D_k\ketbra{0}{0}D_k^{\dagger}$.  
\item The resulting system evolves for a time $\tau$, yielding the transformed state $\rho_k' = \mathcal{E}_{\tau}( \rho_k)$, where $\mathcal{E}_{\tau}$ is given by Eqs.~(\ref{PD_evolution}) or (\ref{AD_evolution}), depending on the type of noise.  
\item  The initial unitary transformation is then reversed, yielding $\rho_k'' = D_k^{\dagger} \rho_k' D_k$.  Note that, in the presence of PDN, if $\rho_k$ had reached one of the target states $\ketbra{e}{e}$ or $\ketbra{g}{g}$, then, after the time evolution, $\rho_k^{\prime}$ would remain in that state and, consequently, $\rho_k''$ would be equal to $\ketbra{0}{0}$.  
In the presence of ADN, this would only hold if  $\rho_k$ had reached the ground state $\ketbra{g}{g}$. 
\item A measurement in the computational basis is performed on the system obtained in the previous step, yielding a result $m_k \in \{0,1\}$. As deduced from the previous discussion, the measurement outcome $m_k=0$ is  compatible with $\rho_k$ having reached one of the target states, while the outcome $m_k=1$ is incompatible with this.
\item Depending on the outcome of $m_k$, the following procedure is applied: 

\noindent $\bullet$ If $m_k = 0$, since the outcome is compatible with having reached one of the target states, a reward is granted by reducing the exploration parameter according to $w_{k+1} = r w_k$, where $ r \in (0,1)$ is a parameter known as the reward rate. Additionally, the unitary transformation $D_k$ remains unchanged, i.e., $D_{k+1} = D_k$,  and the process returns to step $1$.

\noindent $\bullet$ If $m_k = 1$, since the outcome is incompatible with having reached one of the target states, a punishment is applied by increasing the exploration parameter to $w_{k+1}=\mathrm{min}(p w_k,1)$, where $p>1$ is a parameter known as the punishment rate and the $\mathrm{min}$ function ensures that $w_{k+1}$ does not exceed $1$. Additionally, three pseudo-random numbers $\alpha_k$, $\beta_k$, and $\gamma_k$  are generated, uniformly distributed within the exploration interval $[-w_k\pi, w_k \pi]$, and used to construct the pseudo-random rotation
\begin{equation}
R_k=D_k e^{-i \beta_k Y/2}e^{-i \gamma_k Z/2}e^{-i \alpha_k X/2}D_k^{\dagger},
\end{equation} 
where $X=\ketbra{0}{1}+\ketbra{1}{0}$, $Y=-i (\ketbra{0}{1}- \ketbra{1}{0})$, and $Z=\ketbra{0}{0}-\ketbra{1}{1}$ are the Pauli operators. Finally, the unitary transformation $D_k$ is updated to $D_{k+1}=R_k D_k=D_k e^{-i \beta_k Y/2}e^{-i \gamma_k Z/2}e^{-i \alpha_k X/2}$, the qubit is restored to its original state $\ketbra{0}{0}$  by applying the unitary transformation $X$, and the process returns to step $1$.
\end{enumerate}

The previously described algorithm is considered to converge if the exploration parameter $w_k$ approaches zero as $k$ increases. In this case, the pseudo-random rotations $R_k$ tend to the identity operator, and consequently, the unitary transformations $D_k$ converge to a constant value. Moreover, the faster $w_k$ approaches zero, the faster the algorithm converges.

\section{Results}
\label{results}

To examine the impact of the previously discussed noise sources on the algorithm from the preceding section, we have implemented it using a Hamiltonian of the form 
\begin{equation}
\label{Hamiltonian}
H=\frac{\hbar \omega}{4}(\sqrt{3} X-Z),
\end{equation}
which corresponds to setting $\ket{e}=(\ket{0}+\sqrt{3}\ket{1})/2$ and $\ket{g}=(-\sqrt{3}\ket{0}+\ket{1})/2$ in the Hamiltonian expression introduced earlier. To work with dimensionless quantities, we define the parameters $\tilde{\tau}=\omega \tau$ and $\tilde{T}_{\mathrm{D}}=\omega T_{\mathrm{D}}$. The measurement process described in item 4 of the previous section is simulated as follows: at each iteration, we calculate the probability of obtaining $0$ as the measurement outcome using the expression $P_k(0)=\mathrm{Tr}(\ketbra{0}{0} \rho_k'')=\mathrm{Tr}[\ketbra{0}{0}D_k^{\dagger} \mathcal{E}_{\tau}(\rho_k )D_k]=\mathrm{Tr}[\rho_k  \mathcal{E}_{\tau}(\rho_k )]$, with $\mathcal{E}_{\tau}$ given by Eq.~(\ref{PD_evolution}) or (\ref{AD_evolution}), depending on the type of noise considered. Then, a pseudo-random number $\chi_k$ is generated, uniformly distributed in the interval $[0,1]$. If $\chi_k \leq P_k(0)$, the measurement outcome is $m_k = 0$; otherwise, it is $m_k = 1$.

Thanks to the fact that the excited and ground states are known in the considered example, we can use this knowledge to assess the accuracy of the algorithm described in the previous section. To this end, at each iteration, we compute the square root fidelity between the state $\rho_k$ and the stationary states $\ketbra{e}{e}$ and $\ketbra{g}{g}$, given by $f_k^{(e)}=\abs{\mel{e}{D_k}{0}}$ and $f_k^{(g)}=\abs{\mel{g}{D_k}{0}}$, respectively. Since, \textit{a priori}, it is not known which of the two stationary states is closer to $\rho_k$, it is also convenient to consider the highest fidelity between $f_k^{(e)}$ and $f_k^{(g)}$, i.e., $f_k=\max (f_k^{(e)},f_k^{(g)})$. The closer the value of $f_k$ approaches 1 as $k$ increases, the more accurate the algorithm estimation of one of the stationary states will be. 

It is worth mentioning that the quantities $w_k$, $f_k^{(e)}$, $f_k^{(g)}$, and $f_k$ are random variables, whose values will vary from one realization of the algorithm to another. The randomness of these variables arises from two factors: first, the inherent stochastic nature of the measurement outcome $m_k$, and second, the pseudo-random selection of the angles $\alpha_k$, $\beta_k$, and $\gamma_k$. For this reason, it is convenient to perform a large number $N$ of realizations (in our calculations, we use $N=1000$) and consider the arithmetic mean of the values for $w_k$, $f_k^{(e)}$, $f_k^{(g)}$, and $f_k$ obtained in each realization, which will be denoted as $W_k$, $F_k^{(e)}$, $F_k^{(g)}$, and $F_k$, respectively. 

\begin{figure}[t]
	\centering
	\includegraphics[width=1\textwidth]{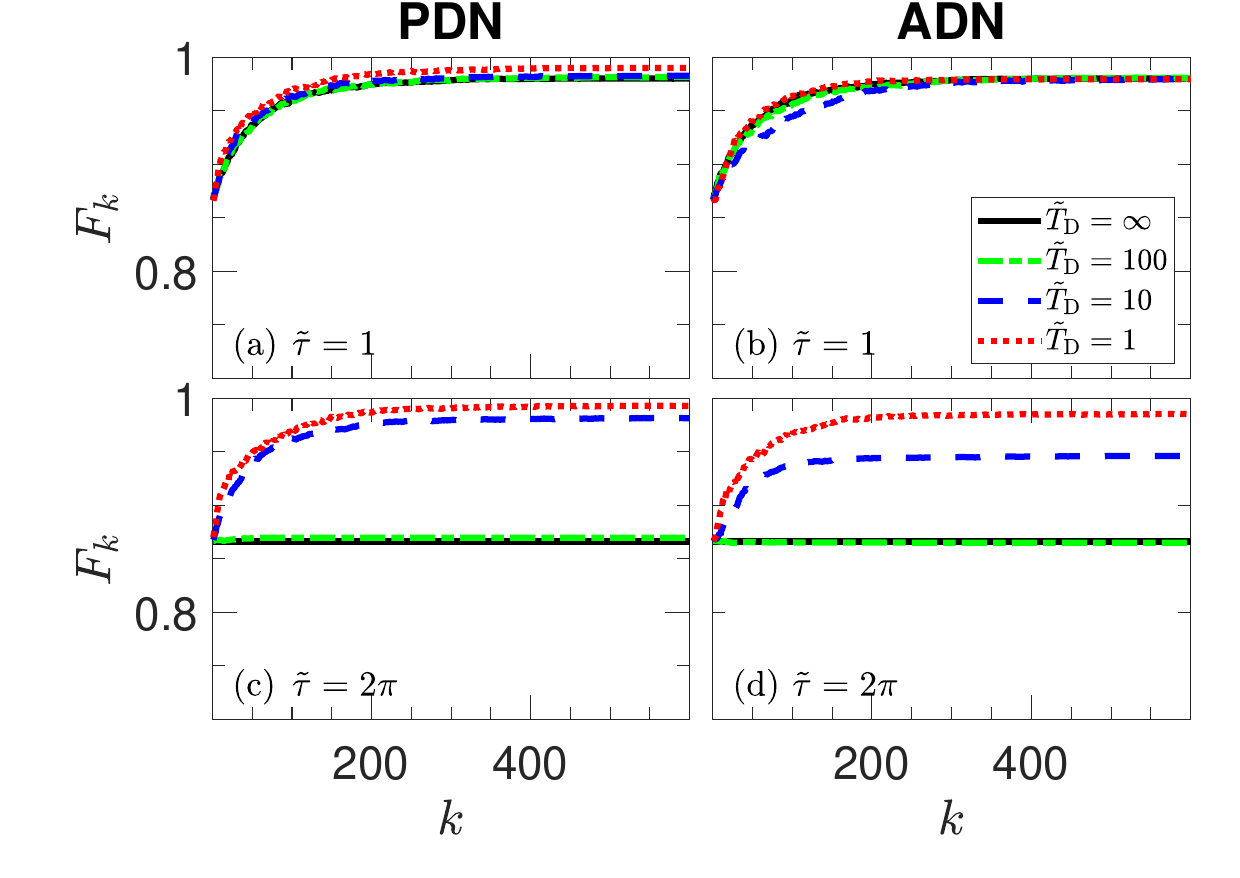}
	\caption{Mean fidelity $F_k$ as a function of the number of iterations $k$ in the presence of PDN (left panels) and ADN (right panels). Results are shown for different dimensionless decoherence times, namely,  $\tilde{T}_{\mathrm{D}} = 1$ (red dotted lines), $\tilde{T}_{\mathrm{D}} = 10$ (blue dashed lines), $\tilde{T}_{\mathrm{D}} = 100$ (green dashed-dotted lines), and $\tilde{T}_{\mathrm{D}} = \infty$ (black solid lines), and for two values of the dimensionless evolution time, specifically,  $\tilde{\tau} = 1$ (top panels) and $\tilde{\tau} = 2\pi$ (bottom panels). The case $\tilde{T}_{\mathrm{D}} = \infty$ corresponds to the scenario with no noise, where the evolution is unitary.}
	\label{Fig1}
\end{figure}

In Fig.~\ref{Fig1}, we depict the mean fidelity $F_k$ as a function of the number of iterations $k$ in the presence of PDN (left panels) and ADN (right panels). The results are shown for several dimensionless decoherence times, namely,  $\tilde{T}_{\mathrm{D}} = 1$ (red dotted lines), $\tilde{T}_{\mathrm{D}} = 10$ (blue dashed lines), $\tilde{T}_{\mathrm{D}} = 100$ (green dashed-dotted lines), and $\tilde{T}_{\mathrm{D}} = \infty$ (black solid lines), and for two values of the dimensionless evolution time, specifically,  $\tilde{\tau} = 1$ (top panels) and $\tilde{\tau} = 2\pi$ (bottom panels). The decoherence time $\tilde{T}_{\mathrm{D}} = \infty$ corresponds to the case without noise, where the evolution is unitary. For all parameter values considered, we have verified the convergence of the algorithm by checking that the mean exploration parameter $W_k$ decreases to zero as the number of iterations $k$ increases sufficiently, although these results are not shown in the figure. As observed in the top left panel, for $\tilde{\tau}=1$, the effect of PDN on the algorithm is minimal, allowing it to perform well even when the decoherence time is comparable to the system's characteristic timescales. In fact, for $\tilde{T}_{\mathrm{D}}=1$, the algorithm performs slightly better than in the absence of noise. In contrast, for $\tilde{\tau} = 2\pi$, the presence of PDN considerably enhances the algorithm performance for dimensionless decoherence times that are not too large, specifically for $\tilde{T}_{\mathrm{D}} = 1$ and $\tilde{T}_{\mathrm{D}} = 10$, with better performance for smaller decoherence times (see bottom left panel). This occurs because, as mentioned in Sec.~\ref{PS},  the unitary evolution operator $U(\tau)$ becomes equal to minus the identity operator for $\tau = 2\pi/\omega$, making all states invariant under the unitary evolution over a time $2\pi/\omega$. As a result, for $\tilde{\tau} = 2\pi$ and in the absence of noise, the algorithm is unable to distinguish between stationary and non-stationary states, causing it to fail. This is not the case in the presence of PDN, in which the stationary states $\ketbra{e}{e}$ and $\ketbra{g}{g}$ are the only pure states that remain invariant under the non-unitary evolution described by Eq.~(\ref{PD_evolution}), even for $\tilde{\tau}=2\pi$. This allows the algorithm to distinguish between stationary and non-stationary states, resulting in a significant improvement in its performance with respect to the noise-free case. In the presence of ADN (right panels), the behavior is consistent with that observed for PDN, and the explanation for the improved performance in the noisy case compared to the noise-free case in the bottom right panel remains applicable. However, in this case, only the stationary state $\ketbra{g}{g}$ remains invariant under the non-unitary time evolution given by Eq.~(\ref{AD_evolution}), unlike in the PDN case, where both stationary states were invariant. 

\begin{figure}[t]
	\centering
	\includegraphics[width=1\textwidth]{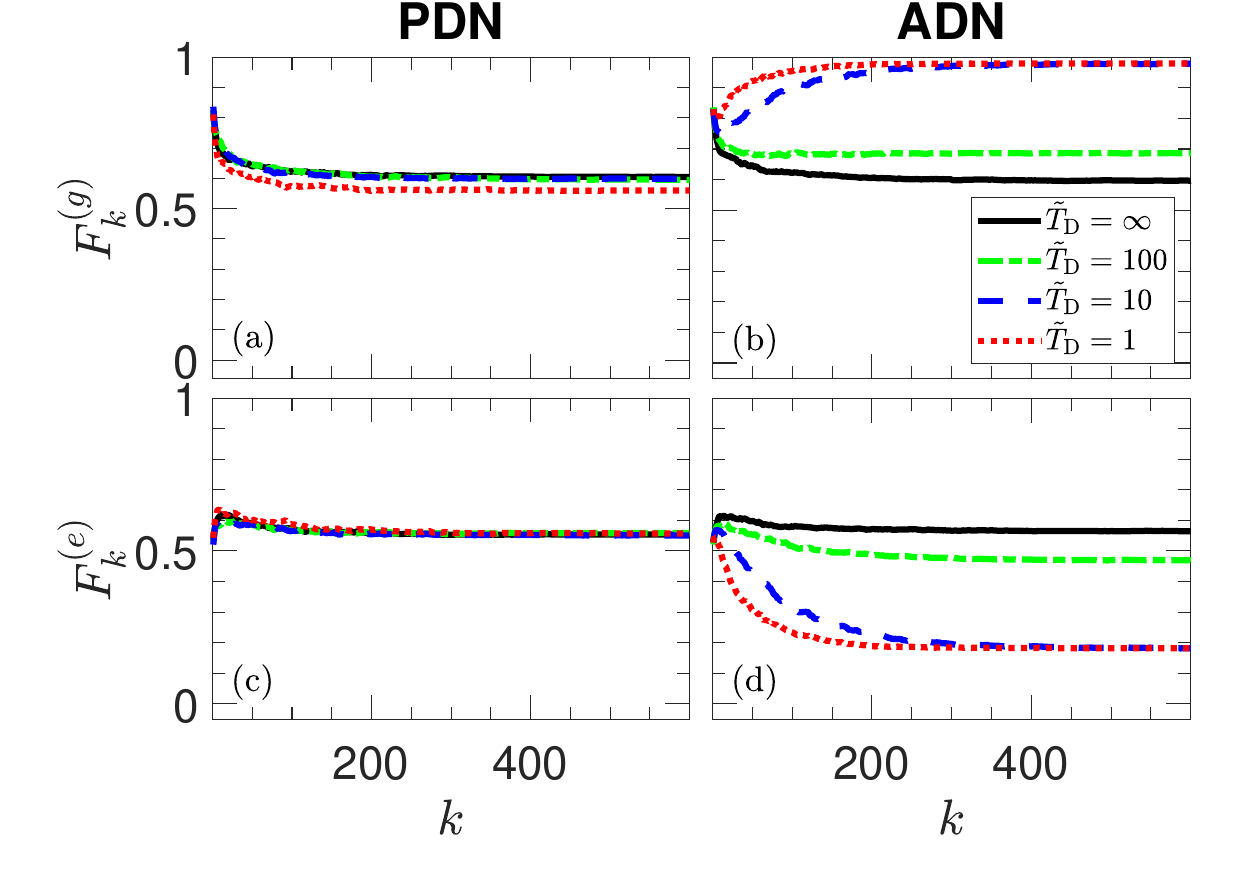}
	\caption{Mean fidelities associated with the ground state, $F_k^{(g)}$ (top panels), and the excited state, $F_k^{(e)}$ (bottom panels), as a function of the number of iterations $k$ for phase damping noise (left panels) and amplitude damping noise (right panels). The values of the dimensionless decoherence times $\tilde{T}_{\mathrm{D}}$ are the same as in Fig.~\ref{Fig1}, and the dimensionless evolution time is $\tilde{\tau} = 1$}
	\label{Fig2}
\end{figure}

To analyze how this difference is reflected in the algorithm performance, Fig.~\ref{Fig2} shows the mean fidelities associated with the ground state, $F_k^{(g)}$ (top panels), and the excited state, $F_k^{(e)}$ (bottom panels), as a function of the number of iterations $k$ for the PDN case (left panels) and the ADN case (right panels).
As seen in this figure, for PDN, the results for $F_k^{(g)}$ and $F_k^{(e)}$ show little dependence on the dimensionless decoherence time $\tilde{T}_{\mathrm{D}}$ and remain close to those obtained in the noise-free case, i.e., for $\tilde{T}_{\mathrm{D}}=\infty$. In contrast, in the presence of ADN, the fidelity $F_k^{(g)}$ increases significantly as $\tilde{T}_{\mathrm{D}}$ decreases, reaching values close to 1 for $\tilde{T}_{\mathrm{D}}=1$ and $\tilde{T}_{\mathrm{D}}=10$, while $F_k^{(e)}$ exhibits a substantial decline. This asymmetry in the behavior of $F_k^{(g)}$ and $F_k^{(e)}$, which is observed for ADN but absent in the PDN case, stems from the distinct properties of the stationary states under each type of noise. As previously mentioned, under PDN, both stationary states, $\ketbra{e}{e}$ and $\ketbra{g}{g}$, remain invariant under the non-unitary time evolution given by Eq.~(\ref{PD_evolution}). As a result, the algorithm converges to either of these states with similar probability, leading to the comparable values of $F_k^{(g)}$ and $F_k^{(e)}$ observed in the left panels of  Fig.~\ref{Fig2}. In contrast, under ADN [Eq.~(\ref{AD_evolution})], only the ground state $\ketbra{g}{g}$ remains invariant. Consequently, the algorithm preferentially converges to this state, leading to a significant enhancement of $F_k^{(g)}$ while $F_k^{(e)}$ decreases accordingly. As the dimensionless decoherence time $\tilde{T}_{\mathrm{D}}$ increases, the non-unitary time evolution (\ref{AD_evolution}) gradually approaches the unitary case $\tilde{T}_{\mathrm{D}}=\infty$, where both $\ketbra{e}{e}$ and $\ketbra{g}{g}$ are once again invariant. This progressively reduces the asymmetry observed for lower values of $\tilde{T}_{\mathrm{D}}$.

\begin{figure}[t]
	\centering
	\includegraphics[width=1\textwidth]{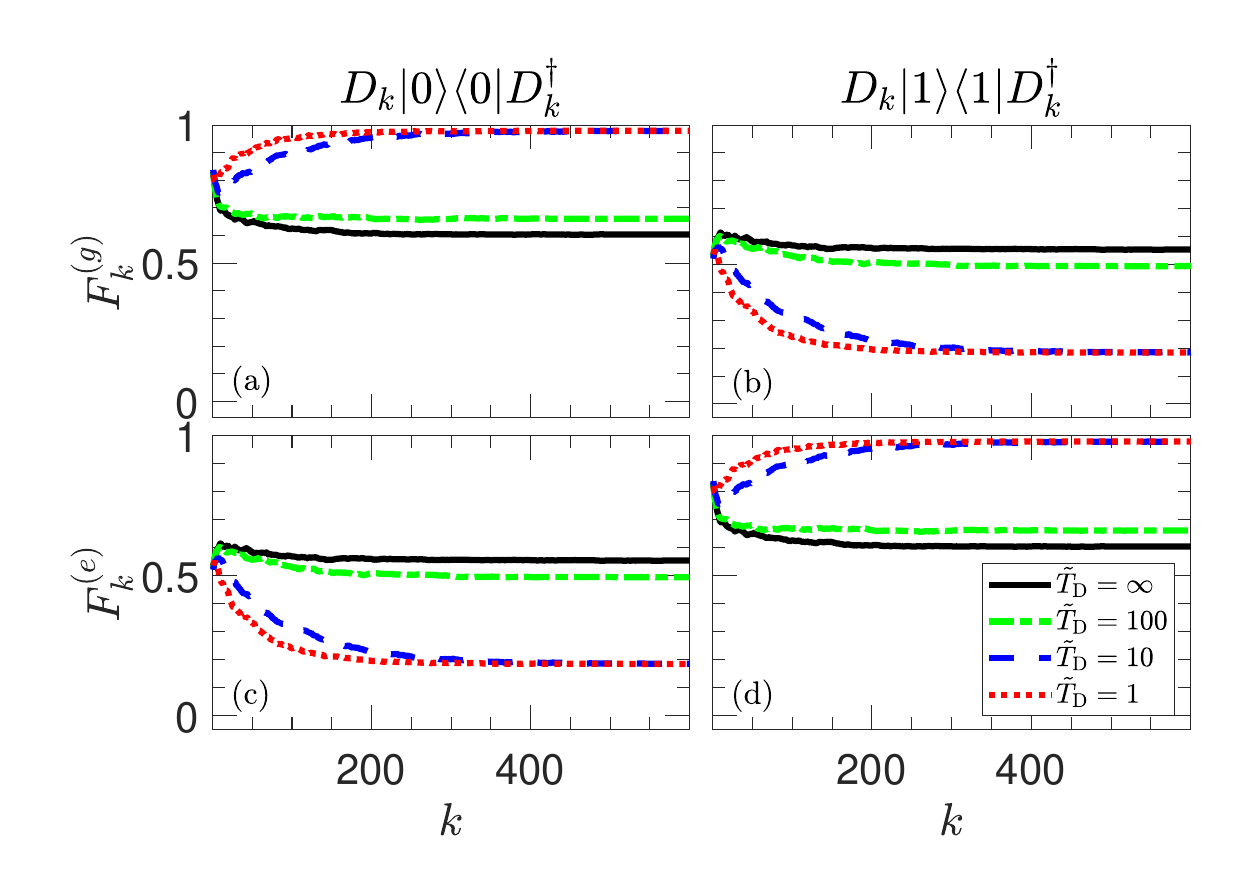}
	\caption{Mean fidelities associated with the ground state, $F_k^{(g)}$ (top panels), and the excited state, $F_k^{(e)}$ (bottom panels), as a function of the number of iterations $k$. The left panels correspond to the states $D_k \ketbra{0}{0} D_k^{\dagger}$, while the right panels correspond to the states $D_k \ketbra{1}{1}D_k^{\dagger}$. The values of $\tilde{T}_{\mathrm{D}}$ and $\tilde{\tau}$ are the same as in Fig.~\ref{Fig2}.}
	\label{Fig3}
\end{figure}

According to the previous discussion, one might conclude that the presence of ADN would be advantageous primarily when aiming to construct the ground state $\ketbra{g}{g}$, while it would be detrimental if the goal were to construct the excited state $\ketbra{e}{e}$. However, the unitary transformations $D_k$ obtained through the previously presented algorithm also allow for an approximate preparation of the excited state by applying them to the computational basis state $\ketbra{1}{1}$ instead of $\ketbra{0}{0}$. Indeed, due to the unitary nature of the operators $D_k$, the state vectors $D_k \ket{1}$ and $D_k \ket{0}$ are orthogonal. Consequently, if $D_k \ketbra{0}{0} D_k^{\dagger}$ is close to the ground state $\ketbra{g}{g}$, then $D_k \ketbra{1}{1} D_k^{\dagger}$ will be close to the excited state $\ketbra{e}{e}$. To confirm this, Fig.~\ref{Fig3} displays the mean fidelities associated with the ground state (top panels) and the excited state (bottom panels) as functions of the number of iterations $k$. The left panels correspond to the states $D_k \ketbra{0}{0}D_k^{\dagger}$, while the right panels correspond to the states $D_k \ketbra{1}{1}D_k^{\dagger}$. The values of $\tilde{T}_{\mathrm{D}}$ and $\tilde{\tau}$ are the same as in Fig.~\ref{Fig3}. As seen in the figure, while the states $D_k \ketbra{0}{0}D_k^{\dagger}$ gradually approach the ground state $\ketbra{g}{g}$ as $k$ increases for $\tilde{T}_{\mathrm{D}}=1$ and $\tilde{T}_{\mathrm{D}}=10$ (top left panel), the states $D_k \ketbra{1}{1}D_k^{\dagger}$  similarly converge to the excited state $\ketbra{e}{e}$ (bottom right panel). In summary, the unitary transformations $D_k$ allow for the calculation of both the ground and excited states by simply applying them to different computational basis states.

\section{Conclusions}\label{conclusions}

In this work, we have analyzed the impact of two common types of noise---PDN and ADN---on the reinforcement learning quantum algorithm proposed in Ref.~\cite{Albarran_Arriagada_2020}. Through the study of specific examples, we have shown that the presence of noise does not necessarily hinder the algorithm performance; in some cases, it can even have a beneficial effect.

In particular, we have demonstrated that for certain values of the evolution time $\tau$,  the presence of noise has little impact when the algorithm accuracy is assessed using the mean fidelity $F_k$. However, for other values of $\tau$, noise can significantly enhance the algorithm performance. Furthermore, we have shown that the two types of noise affect the stationary-state fidelities, $F_k^{(g)}$ and $F_k^{(e)}$, in markedly different ways. While PDN influences both fidelities symmetrically, ADN introduces an asymmetry, favoring convergence to the ground state over the excited state. We have explained this difference by analyzing the pure states that remain invariant under the non-unitary evolution associated with each type of noise.

Although this asymmetry might suggest that ADN enhances the preparation of the system in the ground state compared to the noise-free case, we have also demonstrated that the unitary transformation generated by the algorithm also enables the preparation of the excited state simply by applying it to a different computational basis state.

In this work, for the sake of clarity and simplicity in the description, we have restricted ourselves to the case of a single qubit. However, in future research, we aim to analyze how the results described here generalize when the number of qubits is increased.

\bmhead{Acknowledgements}

The authors acknowledge project PID2022-136228NB-C22 funded by MCIN/AEI/ 10.13039/501100011033 and by ``ERDF A way of making Europe'', EU.  Furthermore, this work has been partially financially supported by the Ministry of Economic Affairs and Digital Transformation of the Spanish Government through the QUANTUM ENIA project call - Quantum Spain project, and by the European Union through the Recovery, Transformation and Resilience Plan - NextGenerationEU within the framework of the ``Digital Spain 2026 Agenda''. It has also been co-financed by EU, Ministerio de Hacienda y
Funci\'on P\'ublica, FEDER and Junta de Andaluc\'{\i}a (project SOL2024-31833).

\section*{Declarations}

All simulation scripts used to produce the results presented in this paper are available at https://github.com/MLOA25/YQS.
%Data sets generated during the current study are available from the corresponding author on reasonable request. 
%\bibliography{literature}% common bib file
%\input ms.bbl



%\end{document}


\end{document}